\documentclass[useAMS,usenatbib,12pt,epsfig]{article}
\usepackage[english]{babel}
\usepackage{amssymb}
\usepackage{graphicx}
\usepackage{wrapfig}
\usepackage{epsf}

\begin{document}
\selectlanguage{english}

\thispagestyle{empty}

\bigskip
\bigskip
\bigskip
\begin{center}
\begin{Large}
{The Thermal Sunyaev-Zeldovich Effect 
\\of Primordial Recombination Radiation}
\end{Large}
\end{center}
\bigskip
\bigskip
\bigskip
\begin{center}
{Kholupenko E.E.$^{\rm 1}$, Balashev S.A.$^{\rm 1,2}$,
Ivanchik A.V.$^{\rm 1,2}$, Varshalovich D.A.$^{\rm 1,2}$}
\end{center}
{$^{\rm 1}$Ioffe Physical-Technical Institute, St.-Petersburg 194021, Russia}
\\{$^{\rm 2}$ St.-Petersburg State Polytechnical University, St.-Petersburg 195251, Russia}
\bigskip
\bigskip
\bigskip
\bigskip
\begin{center}
\bf{Abstract}
\end{center}
{
It is well known that recombination radiation of primordial hydrogen-helium plasma leads to the distortions of 
the planckian spectrum shape of the cosmic microwave background radiation (CMB). We discuss the thermal 
Sunyaev-Zeldovich (SZ) effect with taking into account primordial recombination radiation (PRR). Since in the 
thermal SZ effect the redistribution of the photons depends on the derivatives of the spectrum, the value of 
relative correction to SZ effect due to PRR significantly higher than relative corrections due to PRR in the initial spectrum.

Calculations of corrections to the thermal SZ effect due to PRR show that depending on the cluster parameters:

1) in the range of frequencies $\nu$ = 0.3 - 700 GHz where the cosmic microwave background (CMB) dominates and spectrum
is very close to the planckian one the relative corrections due to PRR have an order of $10^{-9} - 10^{-6}$ of ``pure'' 
SZ effect (i.e. SZ effect for pure planckian spectrum). The difference of intensities of PRR coming from different directions 
(through intracluster and near intercluster medium) reaches values up to 22 mJy/ster at $\nu\simeq 337$ GHz (maximum in considered range).

2) In the range of frequencies $\nu$ = 700 - 5000 GHz where cosmic infrared background (CIB) becomes significant or even
dominates the relative corrections due to PRR can reach $10^{-8}$ - $10^{-5}$ of ``pure'' SZ effect. Corresponding intensity 
difference reaches values up to 25 mJy/ster (at $\nu\simeq 1700$ GHz).

In addition we suggest a modification of the method of electron gas temperature determination using corrections of the SZ effect due to PRR. 
Such modification allows one to simplify the determination of the cluster electron gas temperature in comparison with known methods.

}
\bigskip


\newpage
\section{Introduction}
\hspace{1.1cm}
Thanks to successful missions such as WMAP and Planck and their facility to
discover and investigate cosmological effects with small amplitudes
the interest to the problems of primordial recombination and
radiation arising during this process increased significantly
in the last decade. Many theoretical investigations have been
performed to improve results (e.g. Grachev \& Dubrovich 2008; Switzer \& Hirata 2008; Hirata 2008)
of earlier works (Zeldovich et al. 1968, Peebles 1968, Dubrovich 1975) and to predict
new results (e.g. Grachev \& Dubrovich 2011, Shaw \& Chluba 2011).
 A few techniques have been suggested to observe primordial recombination radiation [PRR]
(Dubrovich \& Grachev 2004; Rubino-Martin et al. 2008; Sunyaev \& Chluba 2008, 2009).
These techniques are difficult to realize because of problems with
cross-calibration of observation channels at different frequencies.
On the other hand the technique of observation of small deviations
from main spectrum at specific frequency but at the different angles
is developed good enough from experimental point of view (e.g. space experiments Relikt-1, COBE, WMAP, Planck, Hershel)
and continues to improve from theoretical point of view (see e.g. Chluba \& Sunyaev 2008, Pitrou et al. 2010, Gruetjen \& Shellard 2012, Shimon et al. 2012,
Chluba et al. 2012, Challinor 2012 and references therein).
Evident power of methods of CMB small angular fluctuation observations
gives us a hope for further successful development of such methods of experimental cosmology to achieve
sensitivities of about $10^{-8}$ (or even $10^{-10}$ in far future) of the background signal.
One of these methods is the observation of the thermal Sunyaev-Zeldovich (SZ) effect. Classical SZ effect (Sunyaev \& Zeldovich 1969)
is the small ($10^{-5} - 10^{-3}$) change of CMB intensity (decrease in Rayleigh-Jeans range [$<218$ GHz] and increase
in Wien range [$>218$ GHz]) due to comptonization of CMB during its transfer through hot (3 - 15 keV) intracluster medium
(i.e. change of photon energies [increase mainly] due to inverse Compton scattering).
In this paper we consider small corrections to the thermal SZ effect in galaxy clusters due to PRR
and suggest a new method for observing PRR via measurements of these corrections by using anisotropy technique.
A similar method allowing to measure the derivative of PRR spectrum using Doppler effect due to motion of observer relative to CMB rest frame 
will be described in the accompanying paper (Balashev et al. 2014 in preparation).
If mentioned sensitivities will be achieved then such measurements will become an important source of information about
PRR characteristics, cosmological parameters, and parameters of galaxy clusters.

\section{Main equations}
In this paper we have estimated simple difference of parameters of
radiation coming through central region of galaxy cluster and through empty
vicinity around it. We have not investigated detailed angular dependence of SZ effect
within cluster angular area and this allowed us to avoid any modeling of internal
structure (e.g. density profile) of hot intracluster medium.
We also suggested that Maxwell-Juttner distribution for electron gas is valid.
This in turn allowed us to use the following approximations for estimate of considered effect:
1) Kompaneets approximation (e.g. Sunyaev \& Zeldovich 1969) that can be easy
tested and gives us an reliable upper estimate of corrections;
2) Generalized kinetic approach (e.g. Birkinshaw 1999) to obtain more exact estimate
of corrections for different values of temperature of electron gas.

In both cases the method of integral kernel $P(\nu, \nu')$ was used, i.e.
the value of occupation number $\eta_{f} \left(\nu\right)$ of radiation
after interaction with hot electron gas is related with initial occupation
number $\eta_{i} \left(\nu\right)$ by the following equation:
\begin{equation}
\eta_{f} \left(\nu\right)=\int_{0}^{\infty}P(\nu, \nu')
\eta_{i} \left(\nu'\right){d\nu' \over \nu'}
\label{Redistribution_relation}
\end{equation}
Equation (\ref{Redistribution_relation}) shows that integral kernel
$P(\nu, \nu')$ is not redistribution function for elementary
photon-particle interaction but describes a complete solution of the problem of
radiation transfer through intracluster medium, i.e. kernel $P(\nu, \nu')$ is the Green's function of corresponding
radiation transfer equation (e.g. Nagirner et al. 1997).
Thus the kernel $P(\nu, \nu')$ depends on cluster parameters such as temperature, density and size.
For transfer through electron gas the kernel $P$ depends on the following
combination of initial and final frequencies: $s=\ln {\nu /\nu'}$.
Therefore hereinafter we use following notation for the kernel: $P_n (s, T_e, \tau)$,
where subscript $n$ denotes meaning of the kernel (e.g. $n=K$ means Kompaneets
approximation, $n=1$ - single-scattering kernel, etc.), $T_e$ is the electron
temperature, $\tau$ is the optical depth to Compton scattering.

In Kompaneets approximation the following kernel is used (Sunyaev \& Zeldovich 1969)
\begin{equation}
P_K (s, T_e, \tau) =
{1 \over \sqrt{4\pi y_C}}
\exp\left(-{\left(s+3y_C\right)^{2} \over 4y_C}\right)
\label{Kompaneets_kernel}
\end{equation}
where $y_C(T_e, \tau)$ is so-called Compton parameter given by the
following formula (in general case):
\begin{equation}
y_C=\int_{0}^{\tau}{k_B T_e \over m_e c^2}d\tau'
\label{y_C_definition}
\end{equation}
where $k_B$ is the Boltzmann constant, $m_e$ is the electron mass,
$c$ is the speed of light. Compton optical depth $\tau$
is the line-of-sight integral through the observed galaxy cluster
\begin{equation}
\tau = \int \sigma_{T} N_e dl
\end{equation}
where $\sigma_{T}$ is the Thomson cross-section, $N_e$ is the concentration of
electron gas, $l$ is the line-of-sight distance. In approximation of constant temperature profile, the Compton
parameter is $y_C=\tau k_B T_e /(m_e c^2)$. The kernel (\ref{Kompaneets_kernel})
describes solution of simple approximation of kinetic equation for photon diffusion
in frequency space as a result of photon-electron scattering (Kompaneets 1957).

In the general kinetic approach we follow Birkinshaw (1999). This
approach gives more exact results and shows that the kernel in formula
(\ref{Redistribution_relation}) depends on cluster parameters not only
by simple product of temperature and optical depth but even in more complex way.
In this case the kernel $P$ for intensity transfer equation is presented in the following form:
\begin{equation}
P_{tot}\left(s, T_e, \tau\right)=\exp(-\tau)
\left(\delta(s)+\sum_{n=1}^{\infty}{\tau^n \over n!}P_{n}(s, T_e)\right)
\label{general_kernel}
\end{equation}
where $P_{n}(s, T_e)$ is the kernel describing radiation intensity redistribution
as result of $n$ scatterings. Kernel $P_{n}$ can be calculated recursively:
\begin{equation}
P_n(s, T_e)=\int_{-\infty}^{+\infty}P_{n-1}(s-t, T_e)P_{1}(t, T_e)dt
\label{P_n_recursion}
\end{equation}
The single-scattering kernel $P_{1}(s, T_e)$ has been calculated according
to the procedure described in details by Birkinshaw (1999). Note that we used common definition of kernels
$P_{tot}$ and $P_{n}$ that implies convolution of these kernels with intensity (e.g. Birkinshaw 1999).
Thus for including $P_{tot}$ in equation (\ref{Redistribution_relation}) where convolution with the occupation number is used
one should use it with the opposite sign of the first argument, i.e. $P_{tot}\left(-s, T_e, \tau\right)$ (for details see Nozawa et al. 2010)

Formulae (\ref{Redistribution_relation} - \ref{P_n_recursion}) have been used for calculation of different
characteristics of the main Sunyaev - Zeldovich effect and correction to it. During the calculations in the frame
of generilized approach the summation in formula (\ref{general_kernel}) was limited by $n=3$ (i.e. scatterings up to third were
taken into account). This approximation is quite adequate since typical observed Compton optical depth of galaxy clusters are below
than $0.1$.

\section{Cosmological model and initial spectra of different radiation components}
The knowledge of incident radiation spectrum is needed to calculate SZ effect.
In considered frequency range there are three important components:
cosmic microwave background (CMB), cosmic infrared background (CIB), and
primordial recombination radiation (PRR):
\\1) CMB spectrum has been considered
as the planckian spectrum with the temperature $T=T_{\rm 0}(1+z)$.
\\2) Data on CIB spectrum were taken from Kneiske \& Dole (2008)
including non-trivial redshift evolution of the spectrum
(i.e. $\nu = \nu_0 (1+z)$ is not enough to describe evolution
of CIB spectrum) due to radiation sources emitting in CIB range
up to present moment $z=0$. Required data were taken from the following
url\footnote{http://www.desy.de/$\sim$kneiske/DATA/MRF$\_$LL.dat}.
\\3) a) PRR spectrum in the frequency range $\nu =$0.3 - 700 GHz has
been taken from Rubino-Martin et al. (2008). Digital data have been obtained
from the following
url\footnote{http://www.iac.es/galeria/jalberto/pages/research-interests/cosmological-recombination.php}.
\\b) PRR spectrum in the frequency range $\nu=$700 - 5000 GHz
has been calculated using modified numerical code ATLANT
\footnote{ATLANT standard version: http://www.ioffe.ru/astro/QC/CMBR/atlant/atlant.html} (Kholupenko et al. (2011)).

All calculations have been performed in the frame of
standard cosmological model. Corresponding values of cosmological
parameters are indicated in Tab. 1. These values have been taken from Bennett et al. (2003), since we have used
results by Rubino-Martin et al. (2008) obtained for these values of cosmological parameters.
The updated values of cosmological parameters by Hinshaw et al. (2013) and Ade et al. (2013)
are in enough agreement [for our aims] with data by Bennett et al. (2003) and will not change results of this paper significantly.
\begin{table}
\begin{center}
\caption{Parameters of the standard cosmological model}
\begin{tabular}{lll}
  \hline
  Value description & Symbol & Value \\
  \hline
  total matter & $\Omega_{\rm tot}$ & ~~~~1 \\
  (in the units of critical density) & & \\
  non-relativistic matter& $\Omega_{\rm m}=\Omega_{\rm CDM}+\Omega_{\rm b}$ & $0.2678$ \\
  baryonic matter& $\Omega_{\rm b}$ & $0.0444$ \\
  relativistic matter& $\Omega_{\rm rel}=\Omega_{\rm \gamma}+\Omega_{\rm \nu}$ & $ 8.23\cdot 10^{\rm -5}$ \\
  vacuum-like energy& $\Omega_\Lambda$ & $0.7322$\\
  Hubble constant & $H_0$ & 71 km/s/Mpc \\
  radiation temperature & $T_{\rm 0}$ & $2.725$ K \\
  helium mass fraction & $Y$ & $0.24$ \\
  \hline
\end{tabular}
\end{center}
\end{table}

\section{Results and discussions}
The following quantities have been choosen as the characteristics of
the main SZ effect and its corrections:
\\1) Relative change of the occupation number (or intensity)
$d = \left(\eta_{f}/\eta_{i}-1\right)$ (we will refer to this quantity
simply as ``SZ effect'').
\\2) Relative change of SZ effect due to distortion of background (CMB+CIB) by
PRR $\Delta d_{prr}/d_{bg}=\left(d_{tot}/d_{bg}-1\right)$,
where $d_{tot}$ is the SZ effect for total spectrum (CMB+CIB+PRR),
$d_{bg}$ is the SZ effect for spectrum of background (CMB+CIB).
We will refer $d_{bg}$ as ``main SZ effect'' in case of general background
and ``standard SZ effect'' for the range where CMB strongly dominates.
\\3) Absolute difference of observed intensities
$\Delta I = \left(I_{f}-I_{i}\right)$, where $I_{i}$ is the intensity of
radiation coming through empty vicinity near the cluster, $I_{f}$ is the
intensity of radiation coming through intracluster medium.
The quantity $\Delta I$ is required to estimate perspectives of
experimental registration of considered corrections from point of view
of modern telescope sensitivity. Hereinafter we will refer to $\Delta I$ as
angular intensity difference.

\subsection{Correction to SZ effect due to PRR in CMB range}
\begin{wrapfigure}[34]{o}{0.9\textwidth}
\begin{center}
\includegraphics[width=0.9\textwidth, clip]{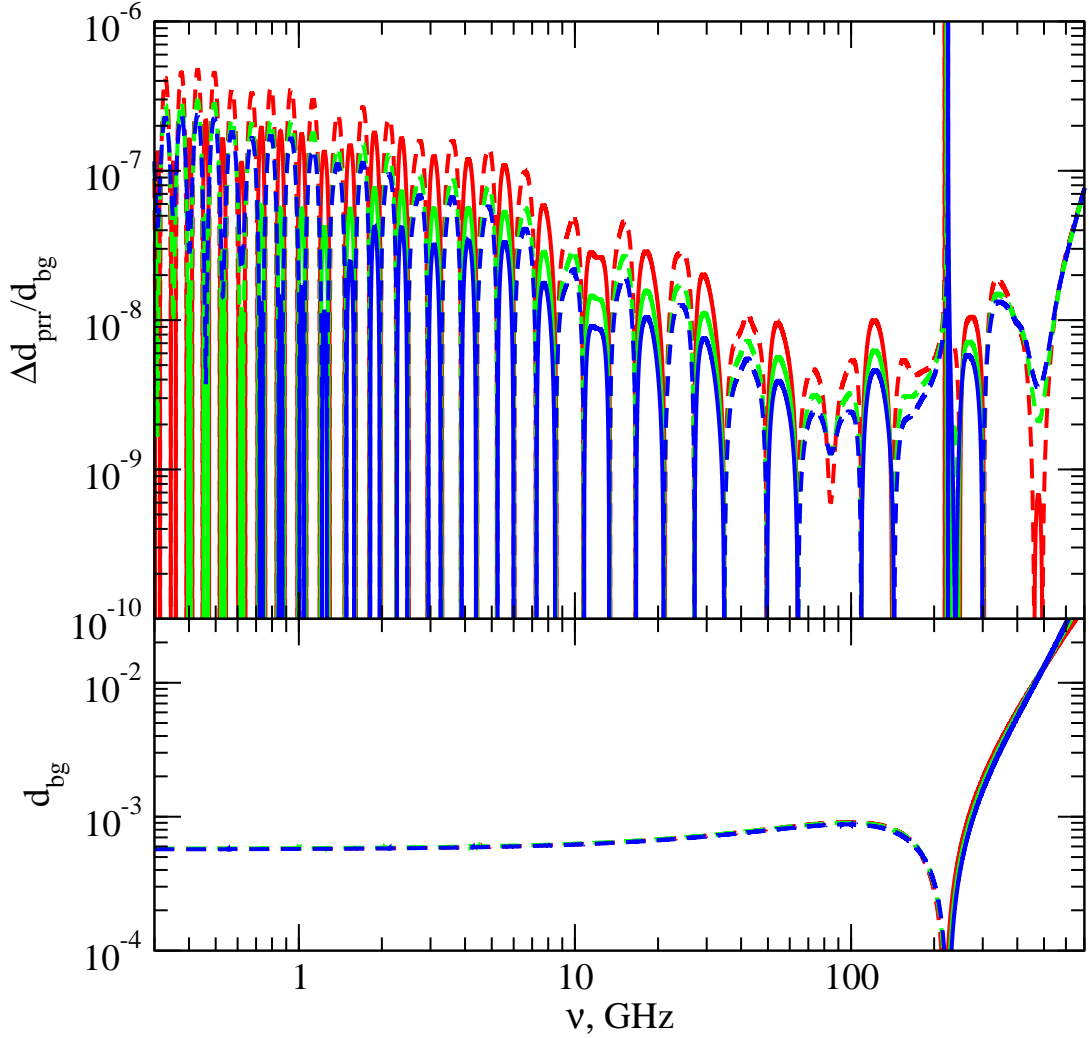}
\end{center}
\caption{{\bf Top panel:} Relative change $\Delta d_{prr}/d_{bg}$ of SZ effect as a function of frequency $\nu$ in the range
$0.3 - 700$ GHz at fixed value of
Compton parameter $y_C=3\cdot 10^{-4}$ and different values of cluster electron gas temperature $T_e$: red, green, and blue
 curves correspond to 5 keV, 10 keV, and 15 keV respectively. Solid curves correspond to the positive part of
$\Delta d_{prr}/d_{bg}$, dashed curves do the negative part of $\Delta d_{prr}/d_{bg}$.
{\bf Bottom panel:} SZ effect $d_{bg}$ as a function of frequency $\nu$.
Curves correspond to the same values of $y_C$ and $T_e$ as in the top panel and are practically overlapped.}
\label{fig1}
\end{wrapfigure}

The results for standard SZ effect $d_{bg}$ at different temperatures $T_e$ of electron gas in the galaxy cluster
and at fixed Compton parameter $y_C=3\cdot 10^{-4}$ are presented in the bottom panel of Fig. \ref{fig1} (it should be noted that
curves corresponding to different temperatures of electron gas are overlapped).
The results for relative change $\Delta d_{prr}/d_{bg}$ of SZ effect at same $T_e$ and $y_C$ values
are presented in the top panel of Fig. \ref{fig1}.
Dependence of $\Delta d_{prr}/d_{bg}$ on observation frequency $\nu$ shows quasi-oscillations.
These oscillations are originated from two phenomena:
\\1) Lines in the initial spectrum (i.e. spectrum of incident PRR).
\\2) Dependence of SZ effect on derivatives (high-orders also) of the initial spectrum. It can be shown by means of Taylor series of initial occupation
number $\eta_{i}$ in expression (\ref{Redistribution_relation}). More evidently and qualitatively this dependence can be shown from
the differential equation of Kompaneets approximation (e.g. Sunyaev \& Zeldovich 1969):
\begin{equation}
{\partial \eta \over \partial y_C}={x_{e}^{-2}}{\partial \over \partial x_{e}} x_{e}^4 {\partial \eta \over \partial x_{e}}
\label{Kompaneets_equation}
\end{equation}
where $x_{e}=h\nu/\left(k_BT_e\right)$ is dimensionless frequency. Note that the kernel (\ref{Kompaneets_kernel}) is the solution
of equation (\ref{Kompaneets_equation}). Let us consider the simple example of the initial spectrum: power spectrum $\eta_{i}\sim x_{e}^{-\alpha}$.
Then SZ effect can be estimated by the following formula: $d=\alpha\left(\alpha-3\right)y_C$ (e. g. Sunyaev \& Zeldovich 1980).
This formula can also be used in the case of slow-changing spectral index $\alpha$  (when $\partial \alpha / \partial \ln x_e \ll \alpha \left(\alpha-3\right)$).
Thus alternating-sign behaviour of value $\Delta d_{prr}/d_{bg}$ can easily be demonstrated by the following relation:
\begin{equation}
\left({\Delta d_{prr} \over d_{bg}}\right)_{K}
\simeq {\eta_{prr}\over \eta_{bg}}{\alpha_{prr}\left(\alpha_{prr}-3\right) \over \alpha_{bg}\left(\alpha_{bg}-3\right)}
\label{D_d_d_analit}
\end{equation}
where subscript ``K'' denotes ``Kompaneets approximation'', $\eta_{bg}$ is the occupation number of background radiation, $\eta_{prr}$ is
the occupation number of PRR spectrum (i.e. the
intitial occupation number is $\eta_{i}=\eta_{bg}+\eta_{prr}$), $\alpha_{bg}\simeq -\partial \ln \eta_{bg} / \partial \ln x_e$,
$\alpha_{prr}\simeq -\partial \ln \eta_{prr} / \partial \ln x_e$. One can see from equation (\ref{D_d_d_analit}) that $\Delta d_{prr} / d_{bg}$ changes its sign
at the points where $\alpha_{prr}\simeq 0$ or 3 and $\alpha_{bg}\simeq 0$ or 3.

Amplitude of these quasi-oscillations
ranges from $2\cdot 10^{-9}$ (at $\nu=100$ GHz) to $5\cdot 10^{-7}$ (at $\nu=0.3$ GHz). Quasi-period $\Delta \nu /\nu$
of quasi-oscillations (i.e. period containing full oscillation) changes from $\simeq 0.1$ (at $\nu=0.3$ GHz) to $\simeq 0.5$ (at $\nu=100$ GHz).
The features at $\nu \simeq 218$ GHz is connected only with the fact that SZ effect $d_{bg}$ goes to zero at this frequency (see bottom panel of Fig. \ref{fig1}).
Fig. \ref{fig1} shows that value $\Delta d_{prr}/d_{bg}$ depends on temperature of electron gas in the galaxy cluster: the hotter the gas, the less
the magnitude of $\Delta d_{prr}/d_{bg}$. This is the result of the kernel shape.
One can see that value $\Delta d_{prr}/d_{bg}$ as a function of frequency $\nu$ has unique shape depending on
parameters of PRR and temperature of electron gas. This should allow researchers to identify reliably observational manifestations of
the effect of thermal corrections to SZ due to PRR.

\begin{wrapfigure}[34]{o}{0.9\textwidth}
\begin{center}
\includegraphics[width=0.9\textwidth, clip]{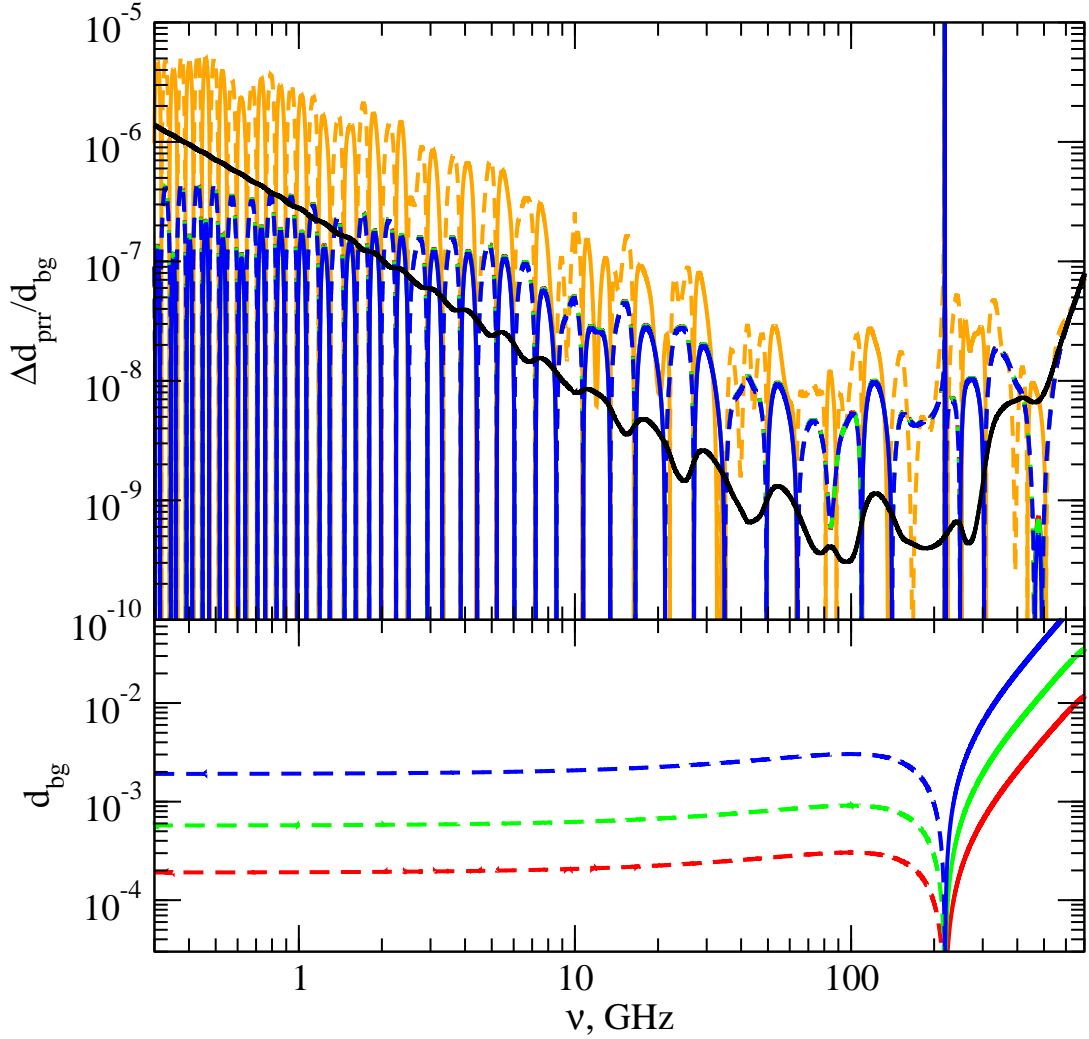}
\end{center}
\caption{{\bf Top panel:} Relative change $\Delta d_{prr}/d_{bg}$ of SZ effect as a function of frequency $\nu$ in the range
$0.3 - 700$ GHz at fixed value of
cluster electron gas temperature $T_e=5$ keV and different values of Compton parameter $y_C$: red, green, and blue curves
correspond to $10^{-4}$, $3\cdot 10^{-4}$, and $10^{-3}$ respectively (these curves are practically overlapped),
orange curves do Kompaneets approximation  at $y_C=3\cdot 10^{-4}$.
Solid curves correspond to the positive part of $\Delta d_{prr}/d_{bg}$, dashed curves do the negative part
of $\Delta d_{prr}/d_{bg}$. Black curve shows the amplitude of PRR relative to background (CMB+CIB, mainly CMB).
{\bf Bottom panel:} SZ effect $d_{bg}$
as a function of frequency $\nu$. Curves correspond to the same values of $y_C$ and $T_e$ as in the top panel.}
\label{fig2}
\end{wrapfigure}

The results of calculations of ``standard'' SZ effect $d_{bg}$ at different values of Compton parameter $y_C$ of the galaxy cluster and at fixed temperature $T_e$
of electron gas are presented in the bottom panel of Fig. \ref{fig2}.
The results of calculations of relative change $\Delta d_{prr}/d_{bg}$ of SZ effect at the same values of $y_C$ and $T_e$ are presented in the top panel of Fig. \ref{fig2}
(it should be noted that curves corresponding to different Compton parameters are overlapped).
Also in the top panel of Fig. \ref{fig2} the relative amplitude of CMB distortions due to PRR is presented.

Here one can emphasize two interesting facts: 1) ``Amplification'' of distortion; 2) Independence of $\Delta d_{prr}/d_{bg}$ on Compton optical depth.

First of them is the following: If we consider CMB intensity $I_{bg}$ as noise (background) $N$, PRR intensity $I_{prr}$ as useful signal $S$,
galaxy cluster as natural filter, SZ effect $d_{bg}$ of CMB as noise $N'$ resulting from pass of noise $N$ through the filter, and correction $\Delta d_{prr}$ to SZ effect
due to PRR as useful signal $S'$ resulting from pass of signal $S$ through the filter, then we would observe an increase (or decrease) of signal-to-noise ratio from
$S/N=I_{prr}/I_{bg}=\eta_{prr}/\eta_{bg}$ to $S'/N' = \Delta d_{prr}/d_{bg}$. In the Kompaneets approximation with slow-changing spectral indecies $\alpha$
corresponding amplification coefficient $C_{K}$ (subscript ``K'' denotes ``Kompaneets'') can be estimated by the following relation resulting from equation (\ref{D_d_d_analit}):
\begin{equation}
C_{K}={S'/N' \over S/N}\simeq{\alpha_{prr}\left(\alpha_{prr}-3\right) \over \alpha_{bg}\left(\alpha_{bg}-3\right)}
\label{C_K}
\end{equation}
One can see from equation (\ref{C_K}) that coefficient of ``amplification'' depends on sharpness of derivative of initial correction spectrum.
For example in the case of correction in Rayleigh-Jeans part of CMB one has $\eta_{bg}\sim {x_e^{-1}}$, and $\alpha_{bg}=1$.
Therefore ``amplification'' coefficient can be rougly estimated as $|C_{K}|\simeq\left|\alpha_{prr}\left(\alpha_{prr}-3\right)/2\right|$.
For $\alpha_{prr}$ achieving values about 8, $|C_{K}|$ achieves values of about 20 (see top panel of Fig. \ref{fig2}).
Exact estimate of the ``amplification'' coefficient $C_{amp}$ by using generilized approach gives us values (for amplitudes, i.e.
at the points of maxima of $|\Delta d_{prr}/ d_{bg}|$) of about $0.25$ - $10$ in Rayleigh-Jeans part of CMB depending on frequency (see top panel of Fig. \ref{fig2}).

Second fact is the following: $\Delta d_{prr}/d_{bg}$ does not depend on Compton optical depth $\tau$ considerably (in the main order).
This is clearly demonstrated by the overlap of curves $\Delta d_{prr}/d_{bg}$ (see top panel of Fig. \ref{fig2}) corresponding to different values of $y_{C}$
at fixed $T_{e}$. It can be shown by using formulae (\ref{Redistribution_relation}) and (\ref{general_kernel}) that the following relation is valid:
\begin{equation}
{\Delta d_{prr} \over d_{bg}}= \left({\eta_{bg}\over \eta_{bg}+\eta_{prr}}
{F\left(T_{e},\left[\eta_{bg}+\eta_{prr}\right]\right) \over F\left(T_{e},\eta_{bg}\right)}-1\right)+O\left(\tau\right)
\label{Delta_d_d_no_tau}
\end{equation}
where the second term (depending on $\tau$) is much less than first term (independing on $\tau$), function $F$ is given by
the following formula:
\begin{equation}
F\left(T_{e},\eta_{i}\right)=\int_{-\infty}^{+\infty}\eta_{i}\left(\nu\exp(-s)\right)P_{1}\left(-s, T_{e}\right)ds
\end{equation}
Formula (\ref{Delta_d_d_no_tau}) shows that the knowledge of $\Delta d_{prr}/d_{bg}$ gives us an opportunity to determine the
temperature $T_{e}$ of electron gas in the galaxy cluster independently of Compton optical depth value $\tau$.
In other words formula (\ref{Delta_d_d_no_tau}) shows how to exclude the dependence of experimantal data (obtained by means of observations of
SZ effect from radiation with non-typical spectra) on Compton optical depth $\tau$ in the main order.
It should be noted that the determination of relative corrections $\Delta d_{prr}/d_{bg}$ is more complicated
than determination of SZ effect $d=\Delta I / I_{i}$ (where both $\Delta I$ and $I_{i}$ are the observable quantities),
because during experiment one can observe only total intensity of radiation (or occupation number
$\left(\eta_{bg}+\eta_{prr}\right)$) and correspondingly can calculate only total SZ effect $\left(d_{bg}+\Delta d_{prr}\right)$, but not ``standard''
SZ effect $d_{bg}$ of background radiation $\eta_{bg}$ (this situation is completely similar to one with possible measurements of PRR lines
distorting CMB spectrum). Thus determination of $\Delta d_{prr} / d_{bg}$  demands special treatment to separate useful signal $\Delta d_{prr}$ and noise signal $d_{bg}$
from the total signal $d_{tot}$. However, this treatment can depend on type of the initial spectrum, features of expected results, quality of data and others.
For example, in the case of determination of $\Delta d_{prr}$ due to PRR, one can average the total SZ effect $d_{tot}$ over frequency on the scales larger than quasi-period of quasi-oscillations
(see Fig. \ref{fig1} and \ref{fig2}) and associate this value with $d_{bg}$.
Also additional assumptions (e.g. assumption on the exact knowledge of CMB spectrum [temperature $T_0$] and its SZ effect) can help us to make a separation of useful and noise signals.

\begin{wrapfigure}[34]{o}{0.9\textwidth}
\begin{center}
\includegraphics[width=0.9\textwidth, clip]{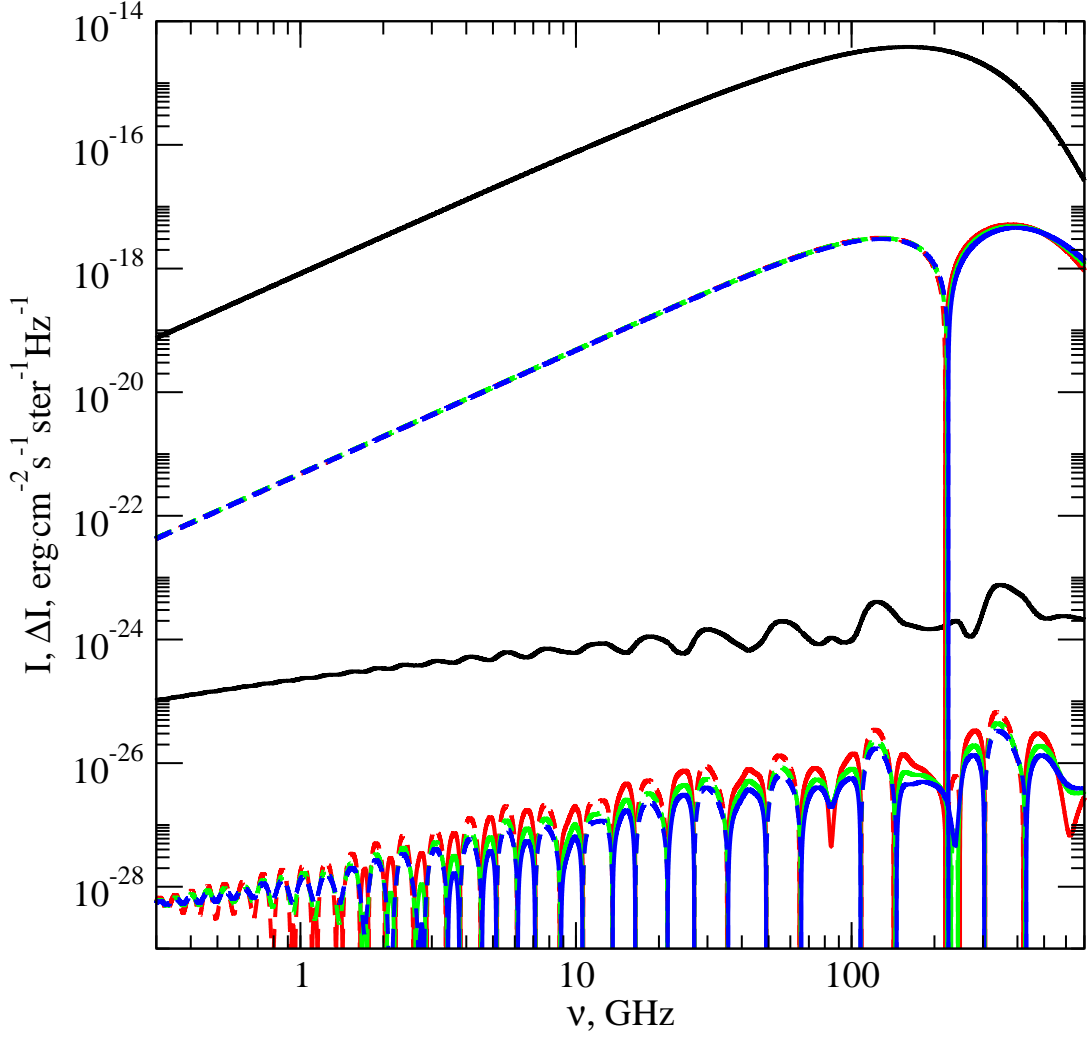}
\end{center}
\caption{Intensities of the initial radiation as functions of frequency $\nu$ in the range
$0.3 - 700$ GHz: the upper black curve corresponds to $I_{bg}$ (background CMB+CIB, mainly CMB),
the lower black curve does $I_{prr}$ (PRR according to Rubino-Martin et al. 2008).  Angular intensity differences
as functions of frequency $\nu$ at fixed Compton parameter $y_C=3\cdot10^{-4}$ for different values of
cluster electron gas temperature $T_e$ are also shown: red, green, and blue
 curves correspond to 5 keV, 10 keV, and 15 keV respectively.
Solid curves correspond to the positive part of $\Delta I$, dashed curves do the negative part of $\Delta I$.
The upper curves (practically overlapped) correspond to angular intensity differences of background radiation (CMB+CIB,
mainly CMB) due to SZ effect, the lower curves do ones of PRR.}
\label{fig3}
\end{wrapfigure}

The results of calculations of intensities $I_{i}$ and their angular differences $\Delta I$ for different initial spectra (background [CMB mainly] and PRR) at different values of
cluster electron gas temperature $T_e$ and at fixed Compton parameter $y_C=3\cdot10^{-4}$ are presented in Fig. \ref{fig3}.
Angular intensity differences for PRR also show marked quasi-oscillations (as well as $\Delta d_{prr}/d_{bg}$).
Amplitude of these quasi-oscillations has values in the range from $6\cdot10^{-29}$ erg$\cdot$cm$^{-2}$s$^{-1}$ster$^{-1}$Hz$^{-1}$ (i.e. $6$ $\mu$Jy/ster at $\nu=0.3$ GHz)
to $6.7\cdot 10^{-26}$ erg$\cdot$cm$^{-2}$s$^{-1}$ster$^{-1}$Hz$^{-1}$ (i.e. $6.7$ mJy/ster at $\nu \simeq 337$ GHz).
Fig. \ref{fig3} also shows that in considered frequency range the larger angular intensity differences for PRR correspond to the lower electron gas temperature while
the angular intensity differences for CMB are almost independent of the temperature of electron gas in the galaxy cluster (at fixed Compton parameter $y_C$).

\begin{wrapfigure}[34]{o}{0.9\textwidth}
\begin{center}
\includegraphics[width=0.9\textwidth, clip]{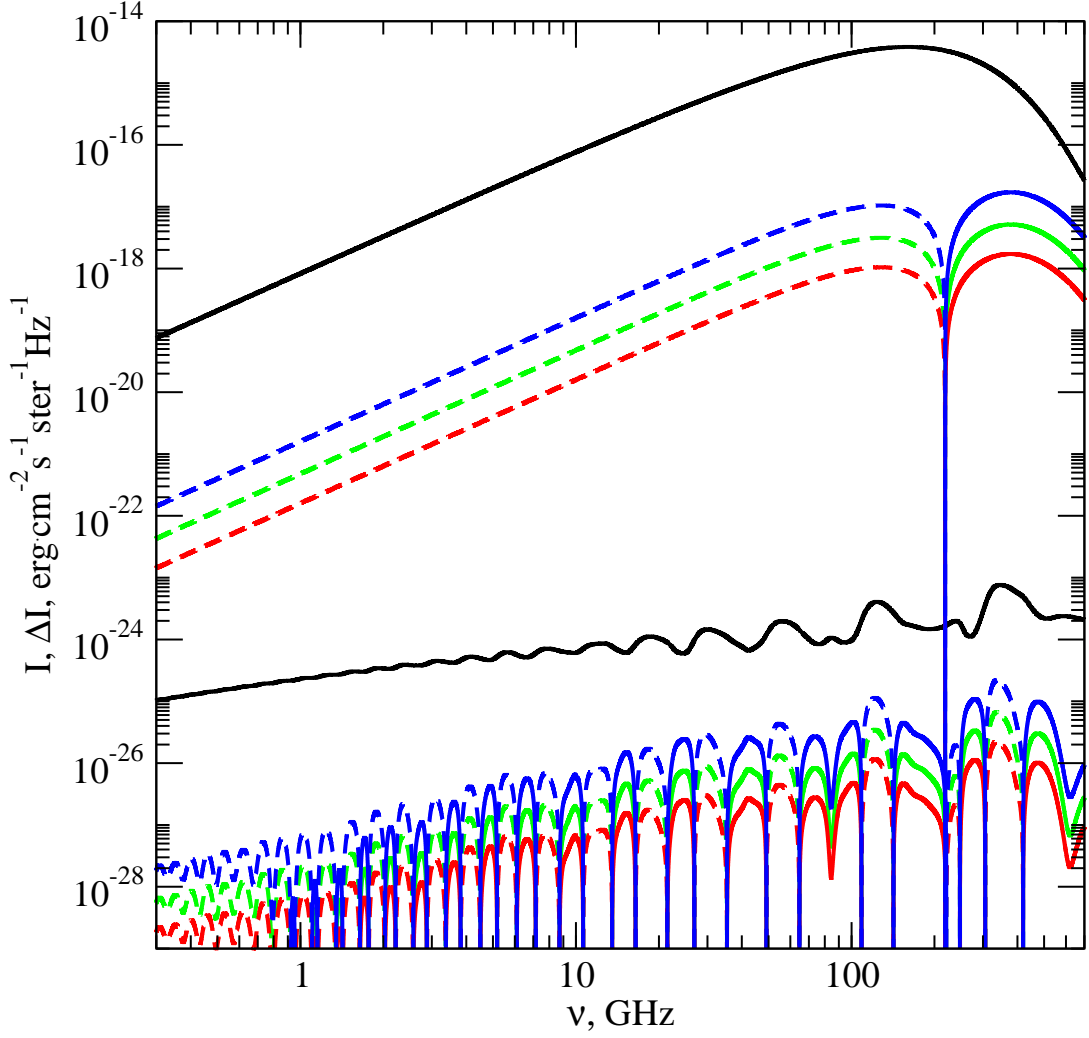}
\end{center}
\caption{Intensities of the initial radiation as functions of frequency $\nu$ in the range
$0.3 - 700$ GHz: the upper black curve corresponds to $I_{bg}$ (background CMB+CIB, mainly CMB),
the lower black curve does $I_{prr}$ (PRR according to Rubino-Martin et al. 2008).  Angular intensity differences
as functions of frequency $\nu$ at fixed cluster electron gas temperature $T_e=5$ keV for different values of
Compton parameter $y_C$ are also shown: red, green, and blue curves correspond to $10^{-4}$, $3\cdot 10^{-4}$,
and $10^{-3}$ respectively. Solid curves correspond to the positive part of $\Delta I$, dashed curves do the negative part of $\Delta I$.
The upper curves correspond to angular intensity differences of background radiation (CMB+CIB, mainly CMB) due to SZ effect,
the lower curves do ones of PRR.}
\label{fig4}
\end{wrapfigure}

The results of calculations of intensities $I_{i}$ and their angular differences $\Delta I$ for different initial spectra (background [CMB mainly] and PRR) at different values of
Compton parameter $y_C$ and at fixed cluster electron gas temperature $T_e=5$ keV are presented in Fig. \ref{fig4}.
One can see that at fixed cluster electron gas temperature $T_e$ angular intensity differences are almost directly proportional to $y_C$, i.e. directly
proportional to Compton optical depth $\tau$ (that is expected for small values of $\tau\ll 1$). For the value of Compton parameter $y_C=10^{-3}$ (at $T_e=5$ keV this
corresponds to $\tau\simeq 0.1$) the amplitude of quasi-oscillations of PRR angular intensity difference achieves about
$2.2\cdot 10^{-25}$ erg$\cdot$cm$^{-2}$s$^{-1}$ster$^{-1}$Hz$^{-1}$ (i.e. $22$ mJy/ster at $\nu \simeq 337$ GHz).


\subsection{Correction to SZ effect due to PRR in CMB-CIB range}
\begin{wrapfigure}[34]{o}{0.9\textwidth}
\begin{center}
\includegraphics[width=0.9\textwidth, clip]{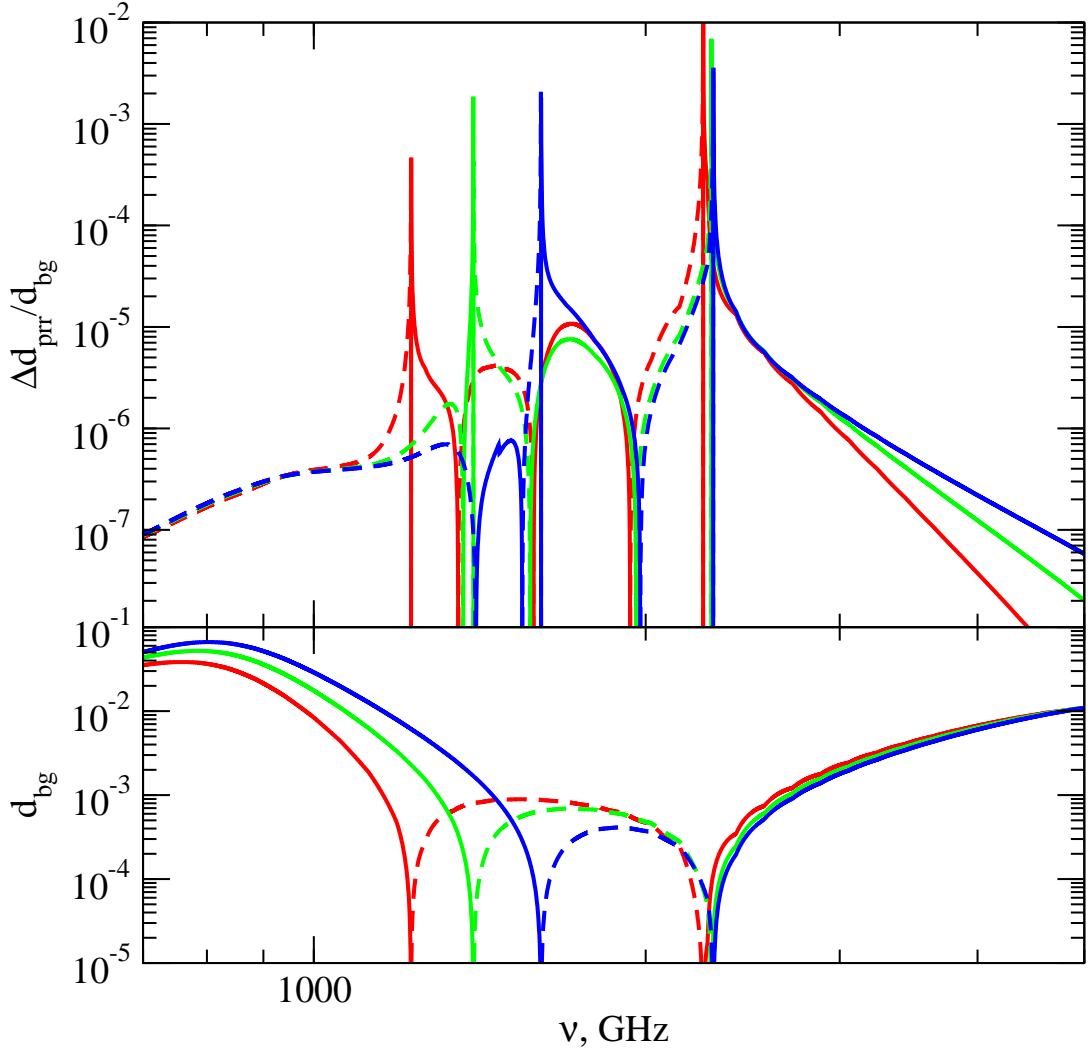}
\end{center}
\caption{{\bf Top panel:} Relative change $\Delta d_{prr}/d_{bg}$ of SZ effect as a function of frequency $\nu$ in the range
$700 - 5000$ GHz at fixed values of Compton parameter $y_C=3\cdot 10^{-4}$ and galaxy cluster redshift $z=0.2$ and
different values of cluster electron gas temperature $T_e$: red, green, and blue
 curves correspond to 5 keV, 10 keV, and 15 keV respectively. Solid curves correspond to the positive part of
$\Delta d_{prr}/d_{bg}$, dashed curves do the negative part of $\Delta d_{prr}/d_{bg}$.
{\bf Bottom panel:} SZ effect $d_{bg}$
as a function of frequency $\nu$. Curves correspond to the same values of $y_C$ and $T_e$ as in the top panel.}
\label{fig5}
\end{wrapfigure}

\begin{wrapfigure}[34]{o}{0.9\textwidth}
\begin{center}
\includegraphics[width=0.9\textwidth, clip]{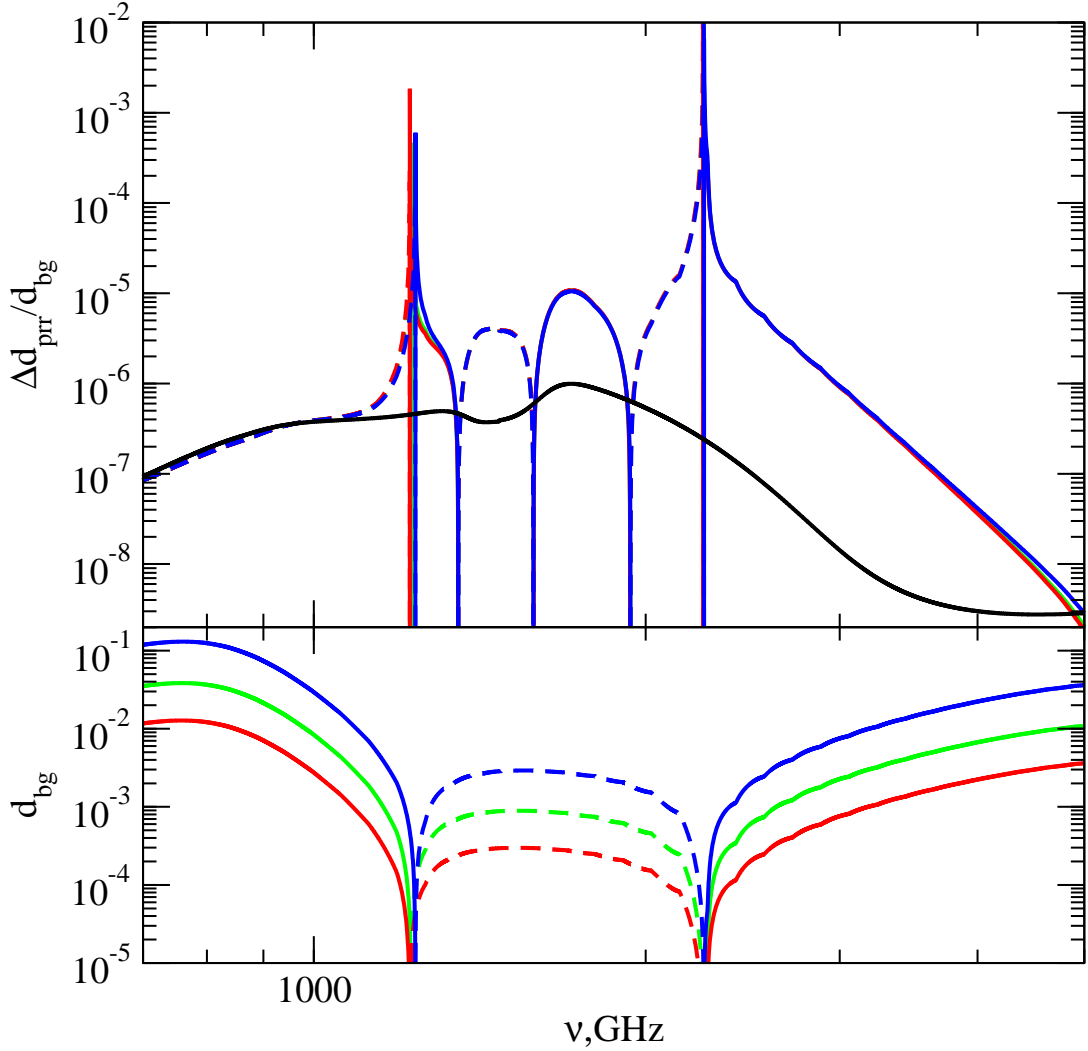}
\end{center}
\caption{{\bf Top panel:} Relative change $\Delta d_{prr}/d_{bg}$ of SZ effect as a function of frequency $\nu$ in the range
$700 - 5000$ GHz at fixed values of cluster electron gas temperature $T_e=5$ keV
and galaxy cluster redshift $z=0.2$ and different values of Compton parameter $y_C$: red, green, and blue curves
correspond to $10^{-4}$, $3\cdot 10^{-4}$, and $10^{-3}$ respectively (these curves are practically overlapped).
Solid curves correspond to the positive part of $\Delta d_{prr}/d_{bg}$, dashed curves do the negative part
of $\Delta d_{prr}/d_{bg}$. Black curve shows the amplitude of PRR relative to background (CMB+CIB).
{\bf Bottom panel:} SZ effect $d_{bg}$
as a function of frequency $\nu$. Curves correspond to the same values of $y_C$ and $T_e$ as in the top panel.}
\label{fig6}
\end{wrapfigure}

\begin{wrapfigure}[34]{o}{0.9\textwidth}
\begin{center}
\includegraphics[width=0.9\textwidth, clip]{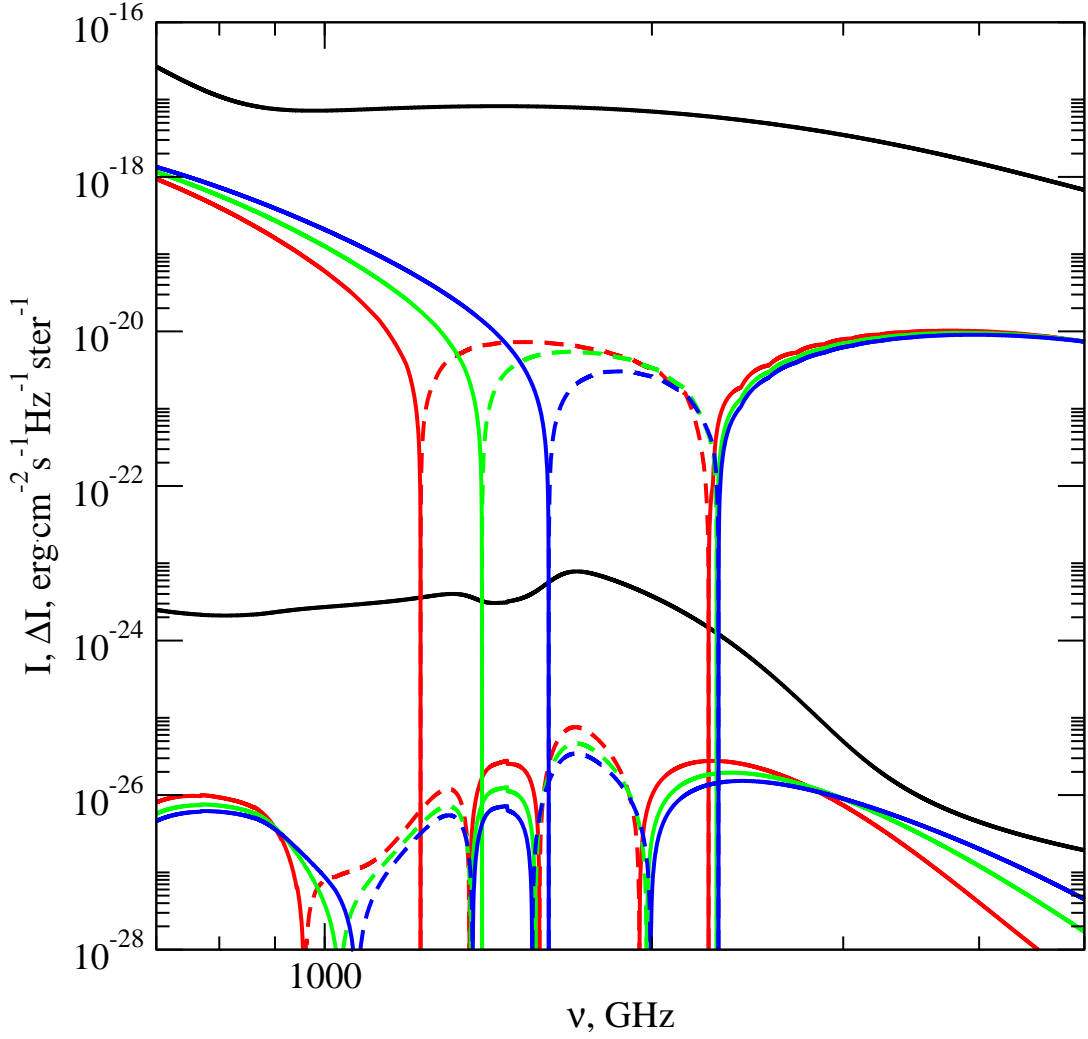}
\end{center}
\caption{Intensities of initial radiation as functions of frequency $\nu$ in the range
$700 - 5000$ GHz: the upper black curve corresponds to $I_{bg}$ (background CMB+CIB),
the lower black curve does $I_{prr}$.  Angular intensity differences
as functions of frequency $\nu$ at fixed Compton parameter $y_C=3\cdot10^{-4}$  and galaxy cluster redshift $z=0.2$
for different values of
cluster electron gas temperature $T_e$ are also shown: red, green, and blue  curves correspond to 5 keV, 10 keV, and 15 keV
respectively. Solid curves correspond to the positive part of $\Delta I$, dashed curves do the negative part of $\Delta I$.
The upper curves correspond to angular intensity differences of background radiation (CMB+CIB) due to SZ effect,
the lower curves do ones of PRR.}
\label{fig7}
\end{wrapfigure}

\begin{wrapfigure}[34]{o}{0.9\textwidth}
\begin{center}
\includegraphics[width=0.9\textwidth, clip]{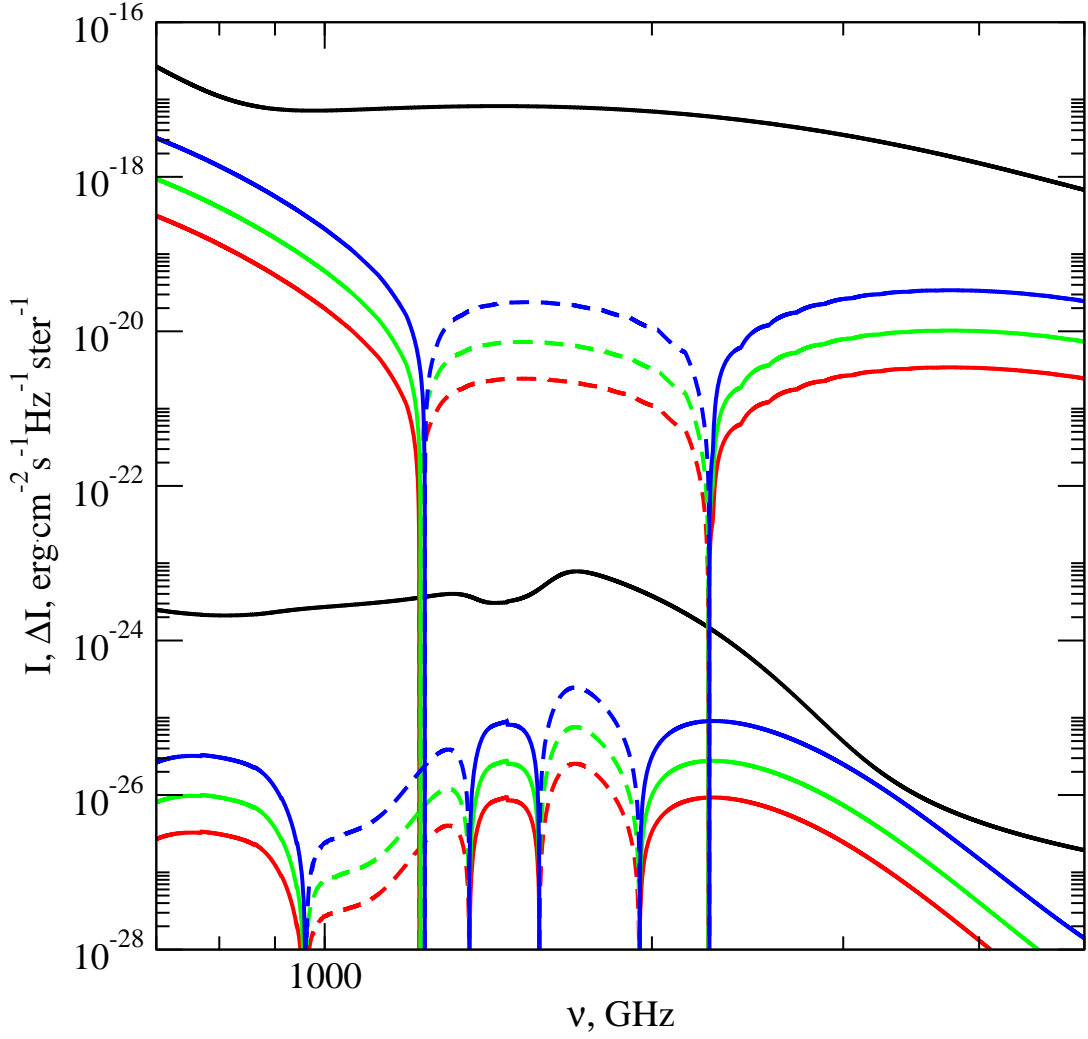}
\end{center}
\caption{Intensities of initial radiation as functions of frequency $\nu$ in the range
$700 - 5000$ GHz: the upper black curve corresponds to $I_{bg}$ (background CMB+CIB),
the lower black curve does $I_{prr}$.  Angular intensity differences as functions of frequency $\nu$
at fixed cluster electron gas temperature $T_e=5$ keV and galaxy cluster redshift $z=0.2$ for different values of
Compton parameter $y_C$ are also shown: red, green, and blue curves correspond to $10^{-4}$, $3\cdot 10^{-4}$,
and $10^{-3}$ respectively. Solid curves correspond to the positive part of $\Delta I$, dashed curves do the negative part of $\Delta I$.
The upper curves correspond to angular intensity differences of background radiation (CMB+CIB) due to SZ effect,
the lower curves do ones of PRR.}
\label{fig8}
\end{wrapfigure}

\begin{wrapfigure}[34]{o}{0.9\textwidth}
\begin{center}
\includegraphics[width=0.9\textwidth, clip]{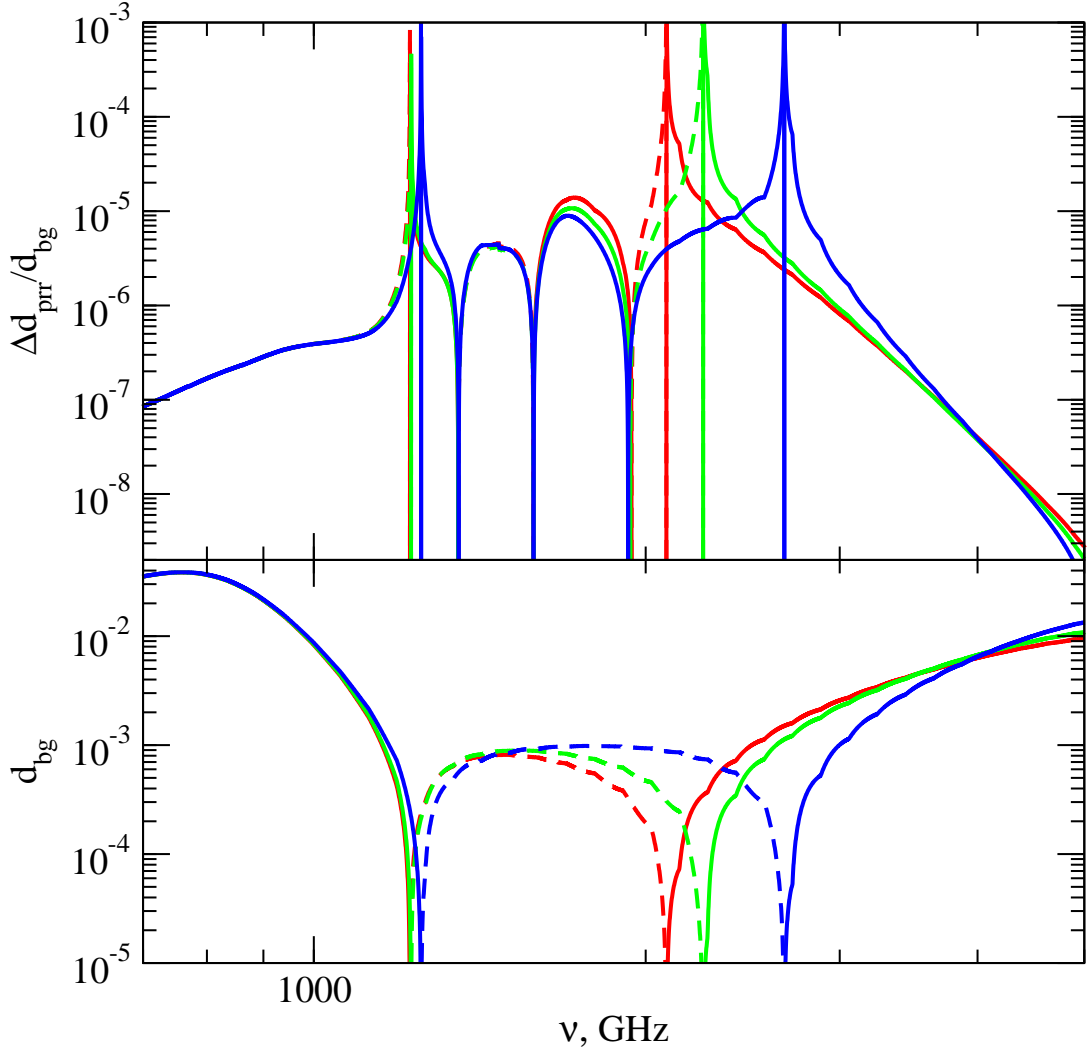}
\end{center}
\caption{{\bf Top panel:} Relative change $\Delta d_{prr}/d_{bg}$ of SZ effect as a function of frequency $\nu$ in the range
$700 - 5000$ GHz at fixed values of Compton parameter $y_C=3\cdot 10^{-4}$ and cluster electron gas temperature $T_e=5$ keV
for different values of redshift $z$ of galaxy cluster: red curves correspond to 0.1, green curves do 0.2, blue curves do 0.5.
Solid curves correspond to the positive part of $\Delta d_{prr}/d_{bg}$, dashed curves do the negative part of $\Delta d_{prr}/d_{bg}$.
{\bf Bottom panel:} SZ effect $d_{bg}$
as a function of frequency $\nu$. Curves correspond to the same values of $y_C$, $T_e$, and $z$ as in the top panel.}
\label{fig9}
\end{wrapfigure}

Calculation results for different characteristics of SZ effect in the frequency range $\nu = 700 - 5000$ GHz are
presented in Figs \ref{fig5} - \ref{fig9}.

It should be noted that CIB spectrum is not well known. We have used the data by Kneiske \& Dole (2008) but the real spectrum of
CIB can differ from used one significantly (it has strong cosmological anisotropy and is liable to the influence of local point sources). Thus the following results on relative
change $\Delta d_{prr}/d_{bg}$ of main SZ effect (see Figs \ref{fig5}, \ref{fig6}) serve only to estimate an order of magnitude of considered effect and
to find the main features of behaviour of $\Delta d_{prr}/d_{bg}$. For precise estimate of $\Delta d_{prr}/d_{bg}$ a new calculation taking specific conditions
(especially the intensity of CIB in specific direction for choosen object) into account should be carried out. For calculations of main SZ effect and correction to it
(Figs. \ref{fig5} - \ref{fig8}) value of redshift of galaxy cluster $z$ has been choosen equal to 0.2.

The results of calculations of SZ effect $d_{bg}$ for background radiation (CMB+CIB) at different temperatures $T_e$ of electron gas in galaxy cluster
and at fixed Compton parameter $y_C=3\cdot 10^{-4}$ are presented in the bottom panel of Fig. \ref{fig5}.
The results of calculations of relative change $\Delta d_{prr}/d_{bg}$ of SZ effect at the same values of $T_e$ and $y_C$
are presented in the top panel of Fig. \ref{fig5}.  The relative change $\Delta d_{prr}/d_{bg}$ has typical values $10^{-8} - 10^{-5}$
except narrow frequency bands where the main SZ effect $d_{bg}$ approaches zero. In these bands the value $\Delta d_{prr}/d_{bg}$ becomes
large (infinite in the points of zeros of main SZ effect) but it is only consequence of definition of $\Delta d_{prr}/d_{bg}$ and does not have any
physical sense. Nevertheless in some bands where main SZ effect has notable values ($d_{bg}\le 10^{-4}$) the relative correction
$\Delta d_{prr} / d_{bg}$ can reach values larger than $10^{-5}$. From Fig. \ref{fig5} one can see that in the difference with the behaviour
of correction $\Delta d_{prr}/d_{bg}$  in the range 0.3 - 700 GHz its behaviour in the range 700 - 5000 GHz depends on gas temperature $T_e$
significantly. The main reason of such dependence is the more complex (from SZ effect point of view) behaviour of background (CMB+CIB) spectrum in the range 700 - 5000 GHz in comparison
with one in the range 0.3 - 700 GHz (almost pure CMB). This leads to appreciable dependence of main SZ effect on temperature $T_e$ (see bottom panel of Fig. \ref{fig5}).
For example frequency of first zero in the range 700 - 5000 GHz depends on $T_e$ approximately as $~T_{e}^{0.2}$. That in turn leads to the significant dependence
of frequency values of poles (and bands of rapid growth) of correction $\Delta d_{prr}/d_{bg}$ on temperature $T_{e}$. Such dependencies of main SZ effect
and its correction on electron gas temperature $T_e$ may be one more useful instruments for determination of $T_e$ and CIB spectrum in the future experiments.

The results of calculations of SZ effect $d_{bg}$ for background radiation (CMB+CIB) at different values of Compton parameter $y_C$
and at fixed temperatures $T_e=5$ keV of electron gas in the galaxy cluster are presented in the bottom panel of Fig. \ref{fig6}.
The results of calculations of relative change $\Delta d_{prr}/d_{bg}$ of SZ effect at the same values of $T_e$ and $y_C$
are presented in the top panel of Fig. \ref{fig6}.  As well as for the correction to the SZ effect $\Delta d_{prr}/d_{bg}$
 in the range 0.3 - 700 GHz one can see from Fig. \ref{fig6} that at fixed $T_{e}$ the main SZ effect $d_{bg}$ depends on Compton parameter $y_{C}$
linearly (i.e. depends on Compton optical depth $\tau$ linearly) while correction $\Delta d_{prr}/d_{bg}$ does not depend on Compton parameter $y_{C}$
in the main order (i.e. does not depend on Compton optical depth $\tau$ in the main order).
In the range 700 - 1100 GHz one can see that $\Delta d_{prr} / d_{bg}$ is practically equal to relative distortion of background spectrum due to PRR.
It is easy to show analitically that in the first order $\Delta d_{prr} / d_{bg} \simeq \left(\Delta I_{prr}/\Delta I_{bg} - I_{prr}/I_{bg}\right)$.
Thus this effect takes place in the case of weak ``amplification'', i.e. when $C_{amp}\ll 1$ (when disturbing spectrum $I_{prr}$ is smooth in comparison with background spectrum $I_{bg}$)
and correspondingly  $\Delta I_{prr}/\Delta I_{bg} \ll I_{prr}/I_{bg}$ and one can write $\Delta d_{prr} / d_{bg} \simeq  - I_{prr}/I_{bg}$.

The results of calculations of intensities $I_{i}$ and their angular differences $\Delta I$ for the different initial spectra (background [CMB + CIB] and PRR) at different values of
cluster electron gas temperature $T_e$ and at fixed Compton parameter $y_C=3\cdot10^{-4}$ are presented in Fig. \ref{fig7}.
As well as in CMB range the angular intensity differences for PRR show marked irregular quasi-oscillations.
Amplitudes of these PRR intensity differences have values in the range from $6.2\cdot10^{-27}$ erg$\cdot$cm$^{-2}$s$^{-1}$ster$^{-1}$Hz$^{-1}$ (i.e. $0.62$ mJy/ster at $\nu \simeq 773$ GHz)
to $7.6\cdot 10^{-26}$ erg$\cdot$cm$^{-2}$s$^{-1}$ster$^{-1}$Hz$^{-1}$ (i.e. $7.6$ mJy/ster at $\nu \simeq 1700$ GHz).
Unlike PRR intensity difference in the range 0.3 - 700 GHz (see Fig. \ref{fig3}) the Fig. \ref{fig7} shows that in the range 700 - 5000 GHz
PRR intensity difference has more complex non-monotonic (in different frequency bands) dependence on electron gas temperature $T_e$.
The background intensity differences also show appreciable dependence on temperature of electron gas $T_e$ (at fixed Compton parameter $y_C$).

The results of calculations of intensities $I_{i}$ and their angular differences $\Delta I$ for the different initial spectra (background [CMB + CIB] and PRR) at different values of
Compton parameter $y_C$ and at fixed cluster electron gas temperature $T_e=5$ keV are presented in Fig. \ref{fig8}.
As well as in the range 0.3 - 700 GHz Fig. \ref{fig8} shows that at fixed cluster electron gas temperature $T_e$ angular intensity differences are almost
directly proportional to $y_C$, i.e. directly proportional to Compton optical depth $\tau$ in the main order.
For the value of Compton parameter $y_C=10^{-3}$ (at $T_e=5$ keV this
corresponds to $\tau\simeq 0.1$) the amplitude of quasi-oscillations of PRR angular intensity difference achieves about
$2.5\cdot 10^{-25}$ erg$\cdot$cm$^{-2}$s$^{-1}$ster$^{-1}$Hz$^{-1}$ (i.e. $25$ mJy/ster at $\nu \simeq 1700$ GHz).

The last picture Fig. \ref{fig9} illustrates behaviour of main SZ effect $d_{bg}$ (the bottom panel) and correction $\Delta d_{prr} / d_{bg}$ at
different values of redshifts of the galaxy cluster. One can see from Fig. \ref{fig9} that at frequencies $\nu$ larger than 1700 GHz
both main SZ effect $d_{bg}$ and correction to it $\Delta d_{prr} / d_{bg}$ show significant dependence on redshift of the galaxy cluster. Such
behaviour differs from the common SZ effect in CMB range. The reason of this difference is non-trivial dependence of CIB spectrum on redshift $z$
due to radiation sources emitting in CIB range during evolution of the Universe, while CMB evolution is described by the simple relation of frequency
redshifting $\nu = \nu_0 (1+z)$. This gives us possibility to determine the redshift of observed galaxy cluster, to separate its proper radiation
in considered range from cosmological background (see also Rephaeli 1995) and to determine contribution to background from IR-sources in
the late epochs ($<z$). Joint analysis of SZ effect and X-ray observations (independent determination of temperature $T_e$ of electron gas) will
allow researchers to find essential limits of these characteristics.

\vspace{0.3cm}
\hspace{0.0cm}
{\bf Acknowledgements}
Authors are grateful to R. A. Sunyaev for careful reading and comments on this manuscript. 
This investigation has been performed with partial support of RFBR grant 13-02-12017-ofi-m 
and grant "Leading Scientific Schools of Russia" NSh-294.2014.2. 
SAB thanks RF Presedent Programme (grant MK-4861.2013.2).

\section{References}
Ade P. A. R., Aghanim N., Armitage-Caplan C., Arnaud M. et  al., 2013, arXiv:1303.5076
\\Balashev S. A., Kholupenko E. E., Ivanchik A. V., Varshalovich D. A., 2014, in preparation
\\Bennett C. L., Halpern M., Hinshaw G., et al. 2003, ApJS, 148, 1
\\Birkinshaw M., 1999, Physics Reports, 310, 97
\\Challinor A., 2012, Proceedings of IAU Symposium S288, 8, S288, 42
\\Chluba J., Nagai D., Sazonov S., and Nelson K., 2012, MNRAS, 426, 510
\\Chluba J., Sunyaev R. A., 2008, A\&A, 488, 861
\\Dubrovich V. K., 1975, Soviet Astron. Lett., 1, 196
\\Dubrovich V. K., Grachev S. I., 2004, Astron. Lett., 30, 657
\\Grachev S. I., Dubrovich V. K., 2008, Astron. Lett., 34, 439
\\Grachev S. I., Dubrovich V. K., 2011, Astron. Lett., 37, 293
\\Gruetjen H. F. and Shellard E. P. S., 2012, arXiv:1212.6945
\\Hinshaw G., Larson D., Komatsu E., Spergel D. N. et al., 2013, ApJ Supplement, 208, id. 19
\\Hirata C. M., 2008, Phys. Rev. D, 78, 023001
\\Kholupenko E. E., Ivanchik A. V., Balashev S. A., Varshalovich D. A., 2011, MNRAS, 417, 2417
\\Kneiske T. M., Dole H., 2008, AIP Conference Proceedings, 1085, 620
\\Kompaneets A., 1957, JETP, 4, 730 (original text in russian ZhETF, 1956)
\\Nagirner D. I., Loskutov V. M., Grachev S. I., 1997, Astrofizika, 40, 349
\\Nozawa S., Kohyama Y., Itoh N., 2010, Phys. Rev. D, 81, 043003
\\Peebles P. J. E., 1968, ApJ, 153, 1
\\Rephaeli Y., 1995, Annual review of astronomy and astrophysics, 33, 541
\\Pitrou C., Bernardeau F., and Uzan J.-P., 2010, JCAP, 07, id. 019
\\Rubino-Martin J. A., Chluba J., Sunyaev R. A., 2008, A\&A, 485, 377
\\Shaw J. R., Chluba J., MNRAS, 415, 1343
\\Shimon M., Sadeh S., and Rephaeli Y., 2012, JCAP, 10, id. 038
\\Sunyaev R. A., Chluba J., 2008, ASP Conference Series, 395, 35
\\Sunyaev R. A., Chluba J., 2009, Astron. Nachr., 330, 657
\\Sunyaev R. A., Zeldovich Ya. B., 1969, Ap\&SS, 4, 301
\\Sunyaev R. A., Zeldovich Ya. B., 1980, Annual review of astronomy and astrophysics, 18, 537
\\Switzer E. R., Hirata C. M., 2008, Phys. Rev. D, 77, 083006
\\Zeldovich Ya. B., Kurt V. G., Syunyaev R. A., 1968, Zh. Eksp. Teor. Fiz., 55, 278
\newpage

\newpage

\newpage

\newpage

\newpage

\newpage

\newpage

\newpage

\newpage
\end{document}